\documentclass[prb,superscriptaddress,twocolumn,floatfix,amsmath,amssymb,showpacs]{revtex4-2}
\usepackage{multirow}%
\usepackage{graphicx}% Include figure files
\usepackage{dcolumn}% Align table columns on decimal point
\usepackage{bm}% bold math
\usepackage{color}
\hfuzz=\maxdimen
\begin{document}

\title{Bulk superconductivity and Pauli paramagnetism in nearly stoichiometric CuCo$_2$S$_4$}

\author{Yu-Ying Jin}
\thanks{These authors contribute equally to this work.}
\affiliation{Department of Physics, Zhejiang Province Key Laboratory of Quantum Technology and Devices, Interdisciplinary Center for Quantum Information, and State Key Lab of Silicon Materials, Zhejiang University, Hangzhou 310027, China}
\author{Shi-Huai Sun}
\thanks{These authors contribute equally to this work.}
\affiliation{Department of Physics, Zhejiang Province Key Laboratory of Quantum Technology and Devices, Interdisciplinary Center for Quantum Information, and State Key Lab of Silicon Materials, Zhejiang University, Hangzhou 310027, China}
\author{Yan-Wei Cui}
\affiliation{Department of Physics, Zhejiang Province Key Laboratory of Quantum Technology and Devices, Interdisciplinary Center for Quantum Information, and State Key Lab of Silicon Materials, Zhejiang University, Hangzhou 310027, China}
\affiliation{School of Science, Westlake Institute for Advanced Study, Westlake University, Hangzhou 310064, China}
\author{Qin-Qing Zhu}
\affiliation{Department of Physics, Zhejiang Province Key Laboratory of Quantum Technology and Devices, Interdisciplinary Center for Quantum Information, and State Key Lab of Silicon Materials, Zhejiang University, Hangzhou 310027, China}
\affiliation{School of Science, Westlake Institute for Advanced Study, Westlake University, Hangzhou 310064, China}
\author{Liang-Wen Ji}
\affiliation{Department of Physics, Zhejiang Province Key Laboratory of Quantum Technology and Devices, Interdisciplinary Center for Quantum Information, and State Key Lab of Silicon Materials, Zhejiang University, Hangzhou 310027, China}
\author{Zhi Ren}
\affiliation{School of Science, Westlake Institute for Advanced Study, Westlake University, Hangzhou 310064, China}
\author{Guang-Han Cao}
\email[corresponding author: ]{ghcao@zju.edu.cn}
\affiliation{Department of Physics, Zhejiang Province Key Laboratory of Quantum Technology and Devices, Interdisciplinary Center for Quantum Information, and State Key Lab of Silicon Materials, Zhejiang University, Hangzhou 310027, China}
\affiliation{Collaborative Innovation Centre of Advanced Microstructures, Nanjing University, Nanjing 210093, China}

\date{\today}% It is always \today, today,
             %  but any date may be explicitly specified

\begin{abstract}
It has long remained elusive whether CuCo$_{2}$S$_{4}$ thiospinel shows bulk superconductivity. Here we clarify the issue by studying on the samples of sulfur-deficient CuCo$_{2}$S$_{3.7}$ and sulfurized CuCo$_{2}$S$_{4}$. The sample CuCo$_{2}$S$_{3.7}$ has a smaller lattice constant of $a=9.454$ {\AA}, and it is not superconducting down to 1.8 K. After a full sulfurization, the $a$ axis of the thiospinel phase increases to 9.475 {\AA}, and the thiospinel becomes nearly stoichiometric CuCo$_{2}$S$_{4}$, although a secondary phase of slightly Cu-doped CoS$_2$ forms. Bulk superconductivity at 4.2 K and Pauli paramagnetism have been demonstrated for the sulfurized CuCo$_{2}$S$_{4}$ by the measurements of electrical resistivity, magnetic susceptibility, and specific heat.
%Possible unconventional superconductivity is discussed.

\end{abstract}

\pacs{74.70.Xa; 72.80.Ga; 61.66.Fn}
%74.70.Xa Pnictides and chalcogenides
%74.10.+v Occurrence, potential candidates
%75.30.Fv Spin-density waves
%61.66.-f Structure of specific crystalline solids (for surface structure, see 68.35.B-)
%61.66.Fn	Inorganic compounds
%72.80.-r	Conductivity of specific materials (for conductivity of metals and alloys, see 72.15.-v)
%72.80.Ga	Transition-metal compounds
%71.45.-d	Collective effects
%71.45.Lr	Charge-density-wave systems (see also 75.30.Fv Spin-density waves)

\maketitle
\section{\label{sec:level1}Introduction}

The discoveries of superconductivity (SC) in the complex copper oxide~\cite{cuprate} and the iron-based pnictide~\cite{FeSC} stimulate enthusiasm to search for SC especially in late 3d-transition-metal (Fe, Co, Ni, and Cu) compounds~\cite{nature2019,Hu2015.PRX,Hu2017.SB,Hu2018.SB,Hu2020.SCP}. Among them, the Co-based superconductors are very limited so far. One example is the cobalt oxyhydrate Na$_{x}$CoO$_{2}\cdot y$H$_{2}$O ($x \approx 0.35, y \approx 1.3$), which shows SC at $T_\mathrm{c}\approx$ 4.5 K~\cite{Takada2003}. The Co-based thiospinel CuCo$_2$S$_4$ shows similarities with Na$_{x}$CoO$_{2}\cdot y$H$_{2}$O in the Co coordination, geometrical frustration, and formal Co valence. However, it is not clear whether CuCo$_{2}$S$_{4}$ superconducts or not. Besides, the magnetism of CuCo$_{2}$S$_{4}$ also remains elusive up to present.

Earlier studies in 1960s suggested Pauli paramagnetism in CuCo$_2$S$_4$~\cite{ssc1967,lotgering1968}, and no superconducting transition was observed down to 0.05 K~\cite{pl1967}. In 1990s, however, it was reported that CuCo$_2$S$_4$ shows a Curie-Weiss (CW) paramagnetism with an effective magnetic moment of 0.89 $\mu_\mathrm{B}$ per formula unit (f.u.)~\cite{Miyatani1993}. A cusp in the magnetic susceptibility appears at $T_\mathrm{N}=$ 18 K, which was attributed to an antiferromagnetic spin ordering. In a multiphasic sample with the nominal composition of Cu$_{1.5}$Co$_{1.5}$S$_4$, SC or superconductor-like behavior was observed with an onset transition temperature of $T_\mathrm{c}^{\mathrm{onset}}=$ 4.0 K and a zero-resistance temperature of $T_\mathrm{c}^{\mathrm{zero}}=$ 2.3 K. Investigations on the $^{63}$Cu and $^{59}$Co NMR suggested a gapless superconducting state as well as antiferromagnetic spin correlations, and SC was considered to be in line with the growth of antiferromagnetic
spin correlation~\cite{nmr.prb1995}. Contrastingly, later NMR study on the Co-rich series samples of (Cu$_{x}$Co$_{1-x}$)Co$_{2}$S$_4$ indicated a full superconducting gap without long-range magnetic ordering for CuCo$_2$S$_4$~\cite{nmr.jpcm2002}. It was concluded that SC and the antiferromagnetic spin correlation are associated with the Co-3d and Cu-3d holes, respectively.

%Notably, the CW paramagnetism was enhanced together with a slightly increased $T_\mathrm{N}$ in Cu$_{1.5}$Co$_{1.5}$S$_4$.
%Miyatani1998.jap
%SC in CuRh$_2$S$_4$ at 4.8 K, but
%CuCo$_2$S$_4$, known as the mineral carrollite, crystallizes in a normal spinel $AB_2X_4$ structure in which Cu and Co respectively occupy the tetrahedral $A$ site and octahedral $B$ site.
%Robbins1967 Hagino1995

One of the present authors (G.-H.C.) and coworkers~\cite{wzl2003} attempted to reproduce SC in CuCo$_2$S$_4$ in 2003. However, no SC was observed above 1.8 K in the single-phase sample of CuCo$_2$S$_4$, although signature of SC at $T_\mathrm{c}=$ 3.5 K was detected in a multiphasic sample. Aito and Sato~\cite{jpsj2004} reported the resistivity data of five CuCo$_2$S$_4$ samples from different batches. Two of them showed a superconducting transition, and the higher $T_\mathrm{c}$ value is 3.8 K. Fang et al.~\cite{whh.prb2005} synthesized an unusual K-doped sample Cu$_{1.3}$K$_{0.2}$Co$_{1.5}$S$_4$ which showed SC at 4.4 K together with a CW-like susceptibility but without any antiferromagnetic transition down to 9 K. A very recent work~\cite{feng2020} showed absence of SC with a weak antiferromagnetic transition at about 4 K in CuCo$_2$S$_4$. In a word, the previous reports on CuCo$_2$S$_4$ appear to be highly dispersive and even contradictive. To our knowledge, evidence of bulk SC with specific-heat measurements has not been reported so far in the Cu-Co-S system.
%It was then stated that a large amount of excess Cu was required for the preparation of the superconducting sample with the actual composition close to CuCo$_2$S$_4$.
%It was also claimed that stoichiometric CuCo$_2$S$_4$ could be synthesized ``with very careful sample preparation procedures'', the sample of which shows a sharp superconducting transition at 4.4 K and a ``perfect Meissner fraction'', but the experimental data were never provided.

The contradictive results above strongly suggest that the physical properties are sensitive to the synthesis of samples, and a controlled preparation of nearly stoichiometric samples of CuCo$_2$S$_4$ is crucial to clarify the intrinsic properties. Previous studies showed difficulties in obtaining desired samples of CuCo$_2$S$_4$~\cite{Miyatani1993,Williamson1974,Craig1975,Craig1979,wzl2003,jpsj2004}. They were commonly synthesized by direct reacting copper and cobalt powders with sulfur in a sealed evacuated silica tube at an elevated temperatures. While relatively low reaction temperatures (500 $^{\circ}$C - 600 $^{\circ}$C) were suggested for the preparation of monophasic CuCo$_2$S$_4$~\cite{Wold1965}, however, a following-up work~\cite{Williamson1974} failed to obtain the single-phase sample. The synthesized sample tends to form CoS$_2$ impurity, as revealed by the phase-relation study in the Cu-Co-S system~\cite{Craig1975,Craig1979}. Indeed, later studies~\cite{Miyatani1993,nmr.jpcm2002,wzl2003,jpsj2004} also showed presence of the CoS$_2$ impurity for the reaction temperatures from 500 $^{\circ}$C to 800 $^{\circ}$C. Note that CoS$_2$ is a ferromagnet with a Curie temperature of $\sim$120 K~\cite{CoS2,CoS2-2008,CoS2-2016}, which makes it more easily to be detected by the magnetic measurement~\cite{wzl2003}.

Here we report a novel two-step strategy for the controlled synthesis of stoichiometric CuCo$_2$S$_4$. First, to minimize the formation of CoS$_2$ impurity, we prepared sulfur-deficient CuCo$_2$S$_{4-\delta}$ with $\delta=$ 0.3. Then, the S-deficient sample was sulfurized by annealing in the presence of appropriate amount of sulfur. As a result, the main phase of the annealed sample was found to be nearly stoichiometric CuCo$_2$S$_4$. Bulk SC at 4.2 K and Pauli paramagnetism in the normal state were demonstrated in the S-compensated CuCo$_2$S$_4$.
%We conclude that sulfur deficiency in CuCo$_2$S$_{4-\delta}$ is the main cause for the absence of SC observed in previous reports.
%The superconducting sample could be synthesized reproducibly through a novel two-step route. In the first step monophasic sulfur-deficient CuCo$_2$S$_{3.7}$ was prepared, which does not show SC. In the second step, then CuCo$_2$S$_{3.7}$ was sulfurized by annealing in the presence of appropriate amount of sulfur. The sulfurized sample, containing nearly stoichiometric CuCo$_2$S$_{4}$ as the main phase, was found to show bulk SC at 4.2 K.
%Pauli paramagnetism is also concluded for CuCo$_2$S$_{4}$.
%sulfur deficiency in CuCo$_2$S$_{4-\delta}$ is detrimental to SC.

%K. Motizuki, Structural Phase Transitions in Layered Transition Metal Compounds, Physics and Chemistry of Materials with A, Vol. 8 (Springer, Netherlands, 2012).

\section{\label{sec:level2}Experimental methods}
%CuCo$_2$S$_{4-\delta}$

Polycrystalline sample of S-deficient CuCo$_2$S$_{3.7}$ was first prepared by high-temperature reactions of the constituent elements in a sealed evacuated silica tube. The source materials were powders of copper (99.997\%), cobalt (99.998\%), and sulfur (99.999\%). The homogenized mixture with the composition of CuCo$_2$S$_{3.7}$ was allowed to fire at 750 $^{\circ}$C for 72 hours. This procedure was repeated to improve the quality of the sample. In the second step, the synthesized CuCo$_2$S$_{3.7}$ was sulfurized in the presence of compensatory sulfur (0.35 S/f.u.) by annealing at 450 $^{\circ}$C for 144 hours in a sealed evacuated silica ampoule. Note that excess of sulfur was necessary to ensure a full sulfurization. This is because, during the sulfurization, a side reaction that forms CoS$_2$ always takes place, which additionally consumes sulfur. Besides, in order to quantize the amount of the CoS$_2$ impurity in the sulfurized sample, we additionally prepared CoS$_2$ by reacting Co with S in an evacuated silica tube. The sample is of single phase with the lattice constant of $a=5.535$ {\AA} (consistent with the previous report~\cite{CoS2}), as determined by the powder x-ray diffractions (XRD).
%We found that the S-deficient sample was monophasic when $\delta \approx$ 0.3.

Powder XRD were carried out using a PANalytical diffractometer (Empyrean Series 2) with a monochromatic Cu-$K_{\alpha1}$ radiation. The crystal structure were refined by a Rietveld analysis using the GSAS+EXPGUI package~\cite{GSAS}. The sulfur content in the crystallites was examined by energy-dispersive x-ray spectroscopy (EDS, Oxford Instruments X-Max) equipped in a scanning electron microscope (SEM, Hitachi S-3700N).

The electrical resistivity and specific heat were measured on a Quantum Design Physical Properties Measurement System, and the magnetic properties were measured on a Quantum Design Magnetic Property Measurement System. The resistivity measurement employed a standard four-terminal method. The heat-capacity measurement utilized a thermal relaxation technique. In the magnetic measurements, the applied magnetic fields were set to be 20 Oe and 10,000 Oe, respectively, to detect SC and to study the normal-state magnetism. In the case of the low-field measurements, both zero-field-cooling and field-cooling protocols were employed.

\section{\label{sec:level3}Results and discussion}

\begin{figure}
\includegraphics[width=8cm]{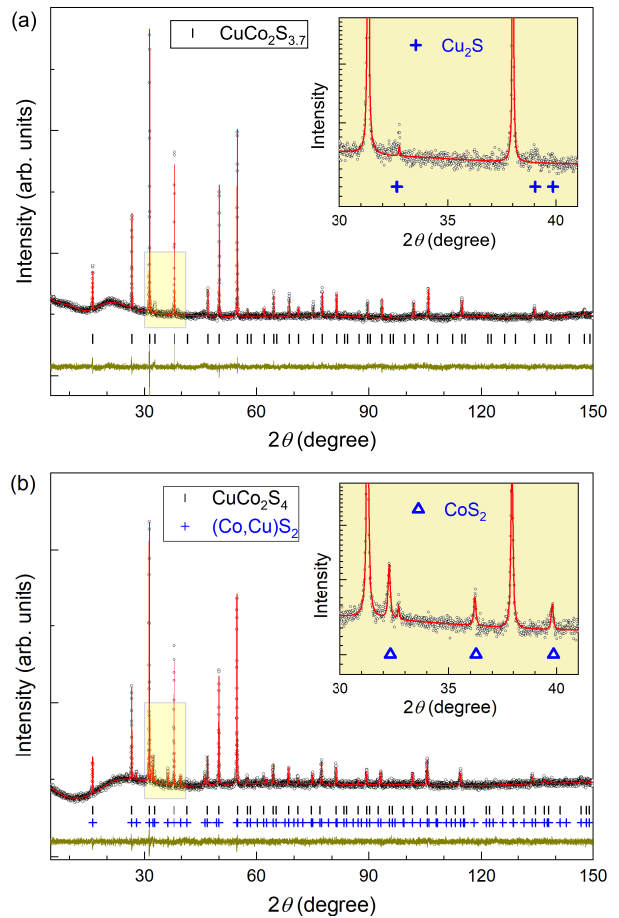}
\caption{Powder X-ray diffractions with the Rietveld refinement profiles for samples of sulfur-deficient CuCo$_2$S$_{3.7}$ (a) and sulfurized CuCo$_2$S$_{4}$ (b). The insets (with a logarithmic scale for the intensity) are a close-up of the marked area, which shows presence of the secondary phases of Cu$_2$S and CoS$_2$, respectively, in CuCo$_2$S$_{3.7}$ and sulfurized CuCo$_2$S$_{4}$.}
\label{xrd}
\end{figure}

Figure~\ref{xrd}(a) shows the XRD profile for the  sulfur-deficient sample of CuCo$_2$S$_{3.7}$. Most of the reflections can be well indexed with a face-centered cubic unit cell of the thiospinel. As is seen in the inset, no reflections associated with CoS$_2$ are detectable, while tiny amount of Cu$_2$S is possibly presented. Therefore, with a lack of sulfur we succeeded in avoiding the appearance of the CoS$_2$ secondary phase. The Rietveld refinement ($R_{\mathrm{wp}}$ = 2.3\% and $\chi^{2}$ = 1.31) confirms the normal spinel structure with $a=9.4544(1)$ {\AA} and $u=0.3865(1)$ for the main phase. Note that the lattice constant is the smallest among those reported previously for CuCo$_2$S$_{4}$ [$9.461(2)$ {\AA}~\cite{Wold1965}, $9.478(5)$ {\AA}~\cite{Williamson1974}, and $9.472(1)$ {\AA}~\cite{Craig1975}]. This may be attributed the apparent sulfur deficiency and/or the partial substitution of Cu by Co (hereafter denoted as Co/Cu substitution)~\cite{nmr.jpcm2002}. The latter is implied by the presence of small amount of Cu$_2$S. As a matter of fact, the Rietveld refinement does not support a significant sulfur vacancy.
%The result suggests that the S-deficient thiospinel is the thermodynamically stable phase at high temperatures.
%So, the sulfur occupancy was fixed to be 0.925 so that the sulfur content remains to be the nominal value, which was confirmed by the EDS measurement to be described below.
%One of the possible reasons is that the sulfur occupancy correlates with the $u$ parameter. We found that the sulfur occupancy decreases with the $u$ parameter, and when $u$ is set to be 3.93, the sulfur occupancy converges to be 0.9375 ($R_{\mathrm{wp}}$ = 2.55\%).

\begin{figure}
\includegraphics[width=8cm]{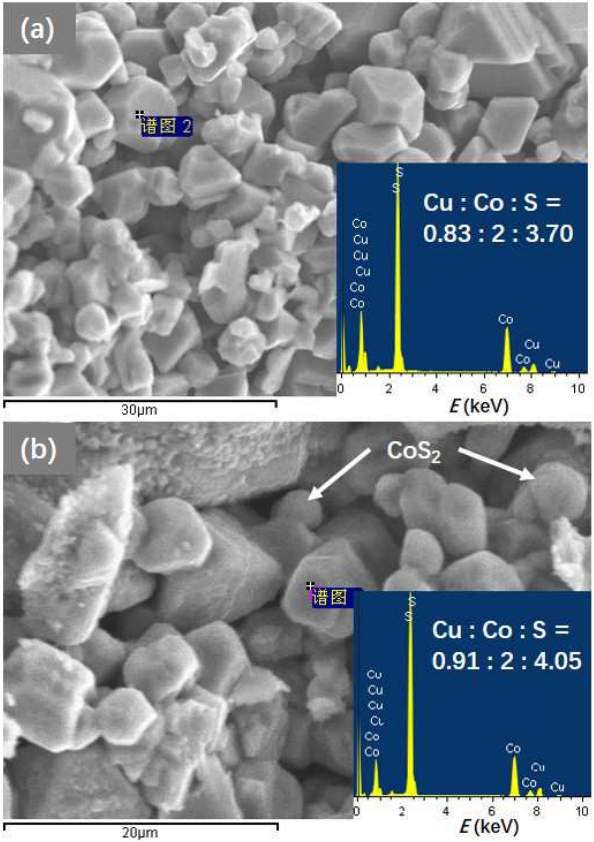}
\caption{Typical scanning electron microscope images of sulfur-deficient CuCo$_2$S$_{3.7}$ (a) and sulfurized CuCo$_2$S$_{4}$ (b). The lower-right insets are the energy-dispersive x-ray spectrum (EDS) collected with the electron beam focused on the spots marked. Round-shape grains (indicated by arrows) can be seen in panel (b), which are identified to be lightly Cu-doped CoS$_2$. The atomic ratios are given by the EDS analysis.}
\label{sem}
\end{figure}

The XRD pattern of the sulfurized CuCo$_2$S$_{4}$ is displayed in Fig.~\ref{xrd}(b). The main phase remains to be the cubic thiospinel, although small amount of CoS$_2$-like phase emerges. With the two-phase Rietveld refinement ($R_{\mathrm{wp}}$ = 2.2\% and $\chi^{2}$ = 1.13), the weight percentage of the CoS$_2$-like impurity was determined to be 14.8(6)\%. The lattice constant of the pyrite phase is refined to be 5.538(1) {\AA}, which is slightly larger than that of CoS$_2$ (5.534 {\AA}~\cite{CoS2}), suggesting that Cu is slightly incorporated. The structural parameters of the main phase were fitted to be $a=9.4750(2)$ {\AA} and $u=0.3851(1)$. The $a$ axis is remarkably larger than that of the sulfur-deficient CuCo$_2$S$_{3.7}$, suggesting a successful sulfurization.
%Note that the $a$ axis reported previously was scattered, ranging from 9.461 {\AA} to 9.482 {\AA}~\cite{Stapele1982.handbook}.

The two samples above were examined by SEM observations in combination with the EDS measurements. As shown in Fig.~\ref{sem}(a), the crystallites of S-deficient CuCo$_2$S$_{3.7}$ are similar in shape. The sulfur content, measured on the basis of the Co content, is consistent with the nominal composition. However, the Cu content is substantially lower than the nominal one. The result suggests that the real composition of the thiospinel phase is something like (Cu$_{1-x-y}$Co$_{x}\Box_{y}$)Co$_2$S$_{4-\delta}$. The SEM image of the sulfurized sample [Fig.~\ref{sem}(b)] shows additional round-shape crystallites which were identified to be slightly Cu-doped CoS$_2$ (1-2\% Cu) by the EDS analysis. Furthermore, the sulfur deficiency is fully compensated, and the Cu content is also increased, as is indicated by the atomic ratio measured. Therefore, we conclude that the sulfurized sample mainly ($\sim$85\% by weight) contains nearly stoichiometric CuCo$_2$S$_{4}$.

\begin{figure}
\includegraphics[width=8cm]{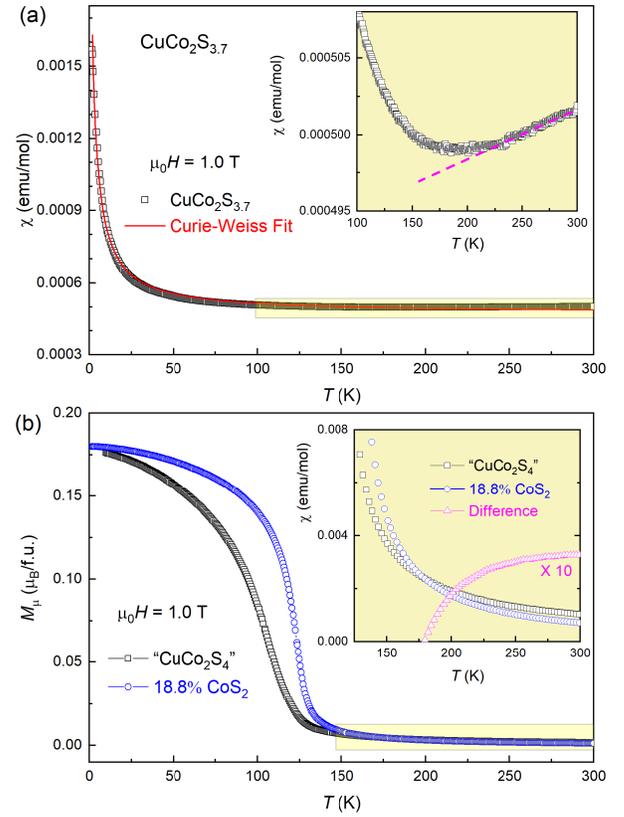}
\caption{Temperature dependence of magnetic susceptibility or magnetization for sulfur-deficient CuCo$_2$S$_{3.7}$ (a) and sulfurized CuCo$_2$S$_{4}$ (b). The inset of (a) is a close-up of the high-temperature data, indicating a positive-temperature-coefficient behavior (dashed line). In panel (b), the magnetization of CoS$_2$ (multiplied by a factor of 18.8\%) is plotted for comparison. The inset of (b) compares the magnetic susceptibilities at high temperatures.}
\label{HFMT}
\end{figure}

Figure~\ref{HFMT}(a) shows the temperature dependence of magnetic susceptibility under a magnetic field of $H=$ 10 kOe for the sulfur-deficient CuCo$_2$S$_{3.7}$. The magnetic susceptibility is nearly temperature independent at high temperatures. No anomaly at $\sim$120 K can be seen, indicating free of the ferromagnetic impurity of CoS$_2$. There is an upturn tail at low temperatures. Fitting of the data with the CW formula, $\chi=\chi_0+C/(T-\theta_\mathrm{CW})$, yields a temperature-independent term of $\chi_0$ = 0.00047 emu mol-f.u.$^{-1}$, a Curie constant of $C$ = 0.0043 emu K mol-f.u.$^{-1}$, and a paramagnetic CW temperature of $\theta_\mathrm{CW} = -1.9$ K. Such a small value of the Curie constant (corresponding to 0.13 $\mu_\mathrm{B}$/Co) is commonly originated from tiny paramagnetic impurities. Additionally, the positive temperature coefficient of susceptibility at high temperatures, shown in the inset of Fig.~\ref{HFMT}(a), also rules out the possible CW-type paramagnetism in CuCo$_2$S$_{3.7}$.

Figure~\ref{HFMT}(b) shows the temperature dependence of magnetization (in the unit of $\mu_\mathrm{B}$/f.u.) of the sulfurized CuCo$_2$S$_{4}$ under the same magnetic field of $H=$ 10 kOe. A ferromagnetic transition is seen at about 120 K, which is attributed to the ferromagnetic impurity of slightly Cu-doped CoS$_2$ that was identified by the XRD experiment above. To quantify the amount of (Co,Cu)S$_2$ independently, the magnetization data of pure CoS$_2$ are shown for comparison. One sees that the Curie temperature of (Co,Cu)S$_2$ is slightly lower than that of CoS$_2$ due to the Cu incorporation. The low-temperature saturation magnetization is about 19\% of that CoS$_2$. At the same time, the high-temperature magnetic susceptibility basically coincides. Since the Cu content in (Co,Cu)S$_2$ is only 1-2\% according to the EDS measurement, the amount of the (Co,Cu)S$_2$ impurity should be also around 19\%, basically consistent with the XRD result above.
% (note that XRD results may be significantly influenced by the phase crystallinity).

The high-temperature magnetic susceptibility data are highlighted in the inset of Fig.~\ref{HFMT}(b), which shows a CW-type paramagnetism. The CW paramagnetism is attributed to the (Co,Cu)S$_2$ impurity, because the magnetic susceptibility of the sulfurized CuCo$_2$S$_{4}$ shows a similar temperature dependence with that of 18.8\% CoS$_2$. The magnetic susceptibility of S-compensated CuCo$_2$S$_{4}$ phase can be roughly obtained by a simple substraction. The result indicates a small value of magnetic susceptibility that is almost temperature independent. Therefore, CuCo$_2$S$_{4}$ should be intrinsically Pauli paramagnetic. Nevertheless, the accurate value of the Pauli-paramagnetic susceptibility cannot be reliably extracted not only because of the influence of the magnetic impurity, but also due to the possible Van Vleck paramagnetism involved~\cite{lotgering1968}. According to the bandstructure calculation of CuCo$_2$S$_{4}$ which gives the density of states at Fermi level of 31.88 states/eV/f.u.~\cite{bandstructure}, the calculated Pauli-paramagnetic susceptibility is derived to be $\chi_\mathrm{P}=\mu_\mathrm{B}^2 N(E_\mathrm{F})=1.03\times10^{-3}$ cm$^3$/mol.
%Anyway, the absence of local-moment magnetism is conforms with monovalence of copper, revealed by soft X-ray photoelectron and absorption spectroscopy in CuCo$_{2}$S$_{4}$~\cite{CuCo2S4.XPS}

\begin{figure}
\includegraphics[width=8cm]{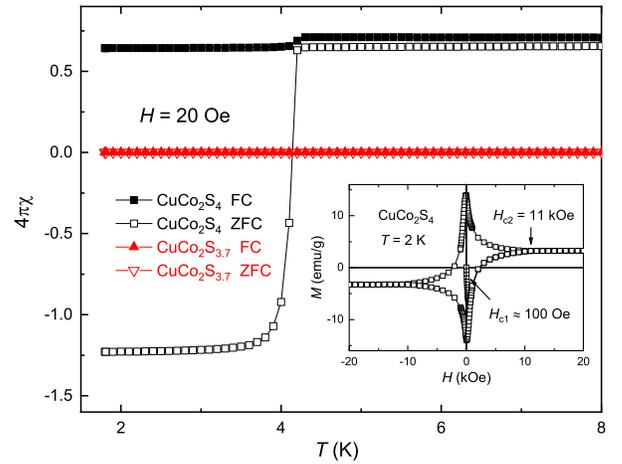}
\caption{Temperature dependence of magnetic susceptibility for sulfur-deficient CuCo$_2$S$_{3.7}$ as well as sulfurized CuCo$_2$S$_{4}$, measured under a magnetic field of 20 Oe in both field-cooling (FC) and zero-field-cooling (ZFC) modes. The inset shows field dependence of magnetization at 2 K for sulfurized CuCo$_2$S$_{4}$.}
\label{MT}
\end{figure}

Figure~\ref{MT} shows the low-temperature susceptibility data for the samples of CuCo$_2$S$_{3.7}$ as well as sulfurized CuCo$_2$S$_{4}$. The S-deficient CuCo$_2$S$_{3.7}$ exhibits low values of magnetic susceptibility, and no signal of SC can be detected down to 1.8 K. By contrast, the sulfurized sample shows a steep decrease in the magnetic susceptibility at 4.2 K, suggesting a superconducting transition. Note that the high value of the susceptibility above $T_\mathrm{c}$ is due to the ferromagnetic impurity (Co,Cu)S$_2$. The large magnitude of the ZFC diamagnetism (exceeding $-100$\%) below $T_\mathrm{c}$ could also be due to the extra magnetic field generated by the ferromagnetic (Co,Cu)S$_2$. The inset shows the field dependence of magnetization at 2 K for the superconducting sample. An extremely type-II superconductivity with $H_{\mathrm{c2}}\gg H_{\mathrm{c1}}$ can be concluded. As expected also, the ferromagnetic signal from (Co,Cu)S$_2$ is superposed on the superconducting loop.

\begin{figure}
\includegraphics[width=8cm]{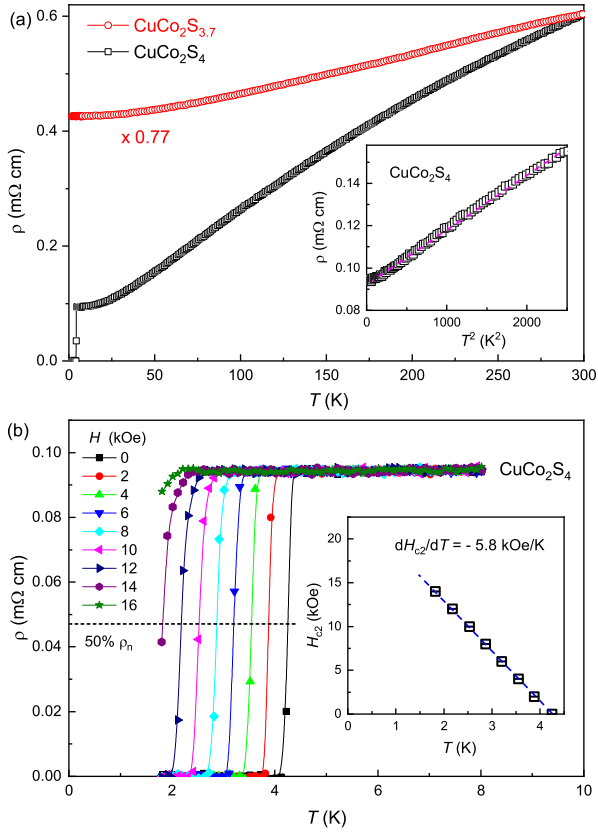}
\caption{(a) Temperature dependence of electrical resistivity ($\rho$) of the polycrystalline samples of sulfur-deficient CuCo$_2$S$_{3.7}$ and sulfurized CuCo$_2$S$_{4}$. The inset plots $\rho$ versus $T^2$ in the temperature range from 4.5 to 50 K. (b) Resistive superconducting transitions under increased magnetic fields from which the upper critical fields $H_{\mathrm{c2}}$ were obtained. The inset plots the resultant $H_{\mathrm{c2}}$ as a function of temperature.}
\label{rt}
\end{figure}

Figure~\ref{rt}(a) shows the temperature dependence of resistivity for the sulfur-deficient CuCo$_2$S$_{3.7}$ and the sulfurized CuCo$_2$S$_{4}$. Both samples show a metallic behavior, yet the sulfurized CuCo$_2$S$_{4}$ sample exhibits a lower room-temperature resistivity with a higher residual resistivity ratio (RRR). The RRR values are 1.4 and 6.4 for CuCo$_2$S$_{3.7}$ and CuCo$_2$S$_{4}$, respectively. Although there are about 19\% (Co,Cu)S$_2$ impurity in the sulfurized CuCo$_2$S$_{4}$ sample, no anomaly at $\sim$120 K associated with the ferromagnetic transition can be detected. At lower temperatures, while no superconducting transition appears down to 1.8 K for CuCo$_2$S$_{3.7}$, a sharp superconducting transition is seen at $T_\mathrm{c}^{\mathrm{onset}}=$ 4.3 K for the S-compensated CuCo$_2$S$_{4}$. The observation of SC in relation with a high RRR value was also reported previously~\cite{jpsj2004}. This could suggest that the nonmagnetic scattering, measured by the residual resistivity, may destroy SC in the system, resembling the scenario in Sr$_2$RuO$_4$~\cite{Ru214} and K$_2$Cr$_3$As$_3$~\cite{K233}. Besides, the low-temperature resistivity of CuCo$_2$S$_{4}$ essentially shows a $T^2$ temperature dependence (see the inset), suggesting dominant electron-electron scattering in the system.

%the subtraction reveals a susceptibility drop of $\sim3\times 10^{-4}$ emu/mol-fu. According to the formula of Pauli-paramagnetic susceptibility, $\chi_\mathrm{P}=\mu_{0}\mu_{\mathrm{B}}^{2}N(E_{\mathrm{F}})$, the susceptibility drop corresponds to a loss of $N(E_\mathrm{F})$ of
The resistive superconducting transitions are more clearly shown in Fig.~\ref{rt}(b). One sees that the superconducting transition shifts to lower temperatures with increasing magnetic fields. Using the criterion of 50\% normal-state resistivity just above $T_\mathrm{c}$ for determining $T_\mathrm{c}(H)$, the upper critical magnetic fields $H_{\mathrm{c2}}$ can be extracted. The resultant $H_{\mathrm{c2}}(T)$ data are shown in the inset of Fig.~\ref{rt}(b), which shows an essentially linear temperature dependence down to 0.42$T_\mathrm{c}$. This result suggests dominant orbital pair-breaking mechanism over a paramagnetic pair-breaking mechanism. The zero-temperature upper critical field is estimated to be $H_{\mathrm{c2}}(0)=$ 24.6 kOe from the linear extrapolation, far below the Pauli-paramagnetic limit $H_\mathrm{P}\approx$ 77 kOe. The coherence length can thus be derived as $\xi_0=$ 11.6 nm using the relation $H_{\mathrm{c2}}(0)=\Phi_0/[2\pi \xi(0)^2]$, where $\Phi_0 (= 2.07\times10^{-15}$ Wb) denotes a magnetic flux quantum.

\begin{figure}
\includegraphics[width=8cm]{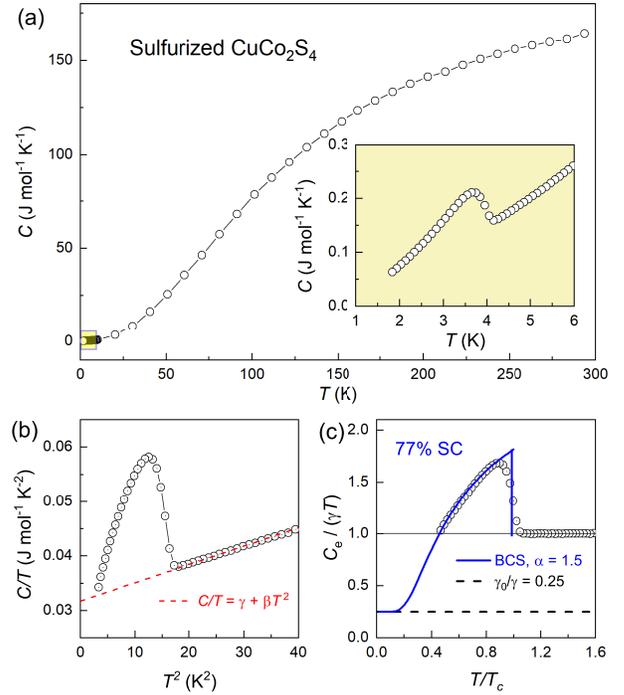}
\caption{Temperature dependence of specific heat for sulfurized CuCo$_2$S$_{4}$. The inset of (a) shows a close-up in the low-temperature region. Panel (b) plots $C/T$ as a function of $T^2$, in which a linear fit ($C/T=\gamma+\beta T^2$) is presented for the normal state. Panel (c) shows $C_\mathrm{e}/(\gamma T)$, where $C_\mathrm{e}=C-\beta T^3$ denotes the electronic specific heat, as a function of the reduced temperature, $T/T_\mathrm{c}$. The data basically agree with a full-gap BCS $\alpha$ model [$\alpha \equiv \Delta(0)/(k_\mathrm{B}T_\mathrm{c})$, where $\Delta(0)$ is the anisotropic superconducting gap at zero temperature]~\cite{alpha} assuming 77\% superconducting phase and a residual electronic specific-heat coefficient of $\gamma_0=0.25 \gamma$.}
\label{sh}
\end{figure}

Figure~\ref{sh}(a) shows the temperature dependence of specific heat for the sulfurized CuCo$_2$S$_{4}$ sample. The specific heat tends to approach the value of $3NR=$ 174.6 J K$^{-1}$ mol$^{-1}$ at high temperatures, in accordance with the Dulong-Petit law. No obvious anomaly is seen at around 120 K where the CoS$_2$ impurity undergoes a ferromagnetic transition. This observation verifies that the CoS$_2$ impurity is the minor phase. As is seen in the inset of Fig.~\ref{sh}(a), at low temperatures, a remarkable specific-heat jump is observable at around 4 K, confirming bulk SC in the sulfurized sample which dominantly contains nearly stoichiometric CuCo$_2$S$_{4}$.
%In fact, the atomic ratio of CoS$_2$ in the sulfurized sample is, 0.19$N_{\mathrm{CoS_2}}/(0.19N_{\mathrm{CoS_2}}+0.81N_{\mathrm{CuCo_2S_{4}}})=$ 9.1\%, where $N_{\mathrm{CoS_2}}$=3 and $N_{\mathrm{CuCo_2S_{4}}}$=7 denote respectively the number of atoms per formula units for the two phases.

Figure~\ref{sh}(b) shows the plot of $C/T$ versus $T^2$, from which the low-temperature electronic specific heat can be separated out. The linear fit gives an intercept of $\gamma$ = 32.2 mJ K$^{-2}$ mol-f.u.$^{-1}$, corresponding to a bare density of states of $N(E_\mathrm{F})=3\gamma/(\pi^2 k_\mathrm{B}^2)=$ 13.6 states/eV/f.u., consistent with the electronic structure calculation~\cite{bandstructure}. Note that the Sommerfeld constant of CoS$_2$ is 21 mJ K$^{-2}$ mol$^{-1}$~\cite{CoS2-2008,CoS2-2016}, somewhat smaller than the above $\gamma$ value, yet it turns out to be larger on the basis of Co content. Furthermore, the CoS$_2$ impurity is the minor phase after all. Therefore, the Sommerfeld coefficient of the CuCo$_2$S$_{4}$ phase will not change very much even if corrections due to the existence of CoS$_2$ impurity could be reliably made. Note that the
%For simplicity and as an approximation, the extracted $\gamma$ value is taken for the CuCo$_2$S$_{4}$ phase.

Assuming the $\gamma$ value of 32.2 mJ K$^{-2}$ mol-f.u.$^{-1}$ and with the electronic specific heat of $C_\mathrm{e}=C-\beta T^3$, figure~\ref{sh}(c) was plotted using $C_\mathrm{e}/(\gamma T)$ and $T/T_\mathrm{c}$ as the coordinates. Under the constrain of entropy conservation, i.e. $\int_0^{T_\mathrm{c}} [(C_\mathrm{e}-\gamma T)/T] \mathrm{d}T=0$, a full-gap BCS $\alpha$ model~\cite{alpha} can basically fit the data with $\alpha \equiv \Delta(0)/(k_\mathrm{B}T_\mathrm{c})=$ 1.5 if a residual electronic specific-heat coefficient of $\gamma_0=0.25 \gamma$ due to the existence of non-superconducting impurity phase of CoS$_2$ is taken into account. In this circumstance, the superconducting fraction is fitted to be 77(1)\%, which is conversely consistent with $\sim$19\% non-superconducting phase.

Here we note that the single-gap BCS model does not account for the data exclusively. Other models with line energy-gap nodes are also applicable. However, the present limited data cannot distinguish which model applies. Interestingly, previous NMR investigations concluded contrasting superconducting properties in the Cu-Co-S system: one suggested a gapless superconducting state~\cite{nmr.prb1995}, the other indicated a full superconducting gap~\cite{nmr.jpcm2002}. This discrepancy seems to be due to the big difference in the sample's quality. Our present specific-heat result excludes the possibility of gapless superconductivity in the nearly stoichiometric sample of CuCo$_2$S$_{4}$. We expect that future measurements of specific-heat, NMR, and other techniques down to lower temperatures with using better samples (with less impurity) will be able to clarify the issue of the superconducting gap.

%The specific heat jump at $T_\mathrm{c}$ is $\Delta C_\mathrm{e}/(\gamma T_\mathrm{c})=$ 1.0 and, if 20\% non-superconducting phase is taken into consideration, the corrected value of $\Delta C_\mathrm{e}/(\gamma T_\mathrm{c})$ will be 1.25.

%Thus, the ideal formula one would expect for a spinel like carrollite is Cu2+Co3+2S24, but as in the case of copper sulfides in general the oxidation state of the copper atom is 1+, not 2+. An assignment of valences as Cu+Co3+2S1.754 is more appropriate; this was confirmed in a study of 2009.[10]

%Electronic environments in carrollite, CuCo2S4, determined by soft X-ray photoelectron and absorption spectroscopy.

Above we have clarified that the nearly stoichiometric CuCo$_2$S$_{4}$ thiospinel is a superconductor with Pauli paramagnetism. Now let us comment on the previous dispersive results about ``CuCo$_2$S$_{4}$''~\cite{pl1967,Miyatani1993,wzl2003,jpsj2004,feng2020}. They can be accounted for in terms of the deviations from the stoichiometry. The actual composition of the synthesized thiospinel phase should be written as (Cu$_{1-x}$Co$_{x}$)Co$_2$S$_{4-\delta}$ (because impurity phases such as CoS$_2$ and Cu$_2$S appeared). The S deficiency obviously decreases the hole concentration in (Cu$_{1-x}$Co$_{x}$)Co$_2$S$_{4-\delta}$, which suppresses SC. The Co/Cu substitution (Co$^{2+}$ partially substitutes Cu$^{+}$) not only decreases the hole concentration, but also possibly induces magnetic impurity of Co$^{2+}$ at the Cu site, both of which are detrimental to SC. This could be the main reason for the difficulty to observe SC in the sample with nominally stoichiometric composition. In the Cu-rich sample of ``Cu$_{1.5}$Co$_{1.5}$S$_4$''~\cite{nmr.prb1995,nmr.jpcm2002}, however, the Co/Cu substitution at the Cu site is greatly reduced because Co is poor. The possible Cu occupation at the Co site may not destroy SC because of nonmagnetic Cu$^+$. Thus SC is easily observed in the Cu-rich samples.

\

\section{\label{sec:level4}Concluding remarks}

To summarize, with a novel two-step synthesis strategy, we were able to prepare nearly stoichiometric CuCo$_2$S$_{4}$ phase which shows bulk SC at 4.2 K with Pauli paramagnetism in the normal state. We have also revealed that sulfur deficiency and Co/Cu substitution is detrimental to SC, which may explain the contradictive results in previous reports. The result calls for further investigations on the rare Co-based superconductor by optimizing the sample quality (with the CoS$_2$ impurity as less as possible) and with various measurements down to lower temperatures.

SC in Co-based compounds is very rare. This work corroborates that CuCo$_2$S$_{4}$ is another Co-based superconductor in addition to Na$_{x}$CoO$_{2}\cdot y$H$_{2}$O~\cite{Takada2003}. Albeit of different crystal structures, interestingly, the two systems show many similarities including the $T_\mathrm{c}$ value, Co coordination, formal Co valence, and the geometrical frustration. It is of great interest to clarify whether CuCo$_2$S$_{4}$ is an unconventional superconductor~\cite{PC2015}. On the other hand, SC is not frequently found in the thiospinel compounds. However, the Cu$M_2$S$_{4}$ ($M=$ Co, Rh, or Ir) family seem to be the only exception. CuRh$_2$S$_{4}$ was first discovered to be a superconductor in 1967 with $T_\mathrm{c}$ = 4.35-4.8 K~\cite{pl1967,Robbins1967}, which was confirmed in 1990s~\cite{Hagino1995}. CuIr$_2$S$_{4}$~\cite{Nagata1994} itself is not a superconductor, yet it undergoes a metal-insulator transition at 230 K accompanied with a charge ordering as well as a spin dimerization~\cite{Radaelli2002}. SC with $T_\mathrm{c}$ up to 3.4 K can be induced by the suppression of the metal-insulator transition via Zn/Cu substitution~\cite{Cao2001,Suzuki1999}. For the spinel selenides, SC was reported in CuRh$_2$Se$_{4}$ ($T_\mathrm{c}=$ 3.5 K~\cite{pl1967,Robbins1967}) and Cu(Ir$_{0.8}$Pt$_{0.2}$)$_2$Se$_{4}$ ($T_\mathrm{c}=$ 1.76 K~\cite{Cava2013}). Therefore, one may expect that CuCo$_2$Se$_{4}$ could be also a superconductor if it can be synthesized with the stoichiometric composition.

\begin{acknowledgments}
This work was supported by the National Natural Science Foundation of China (12050003), National Key Research and Development Program of China (2017YFA0303002) and the Fundamental Research Funds for the Central Universities of China.

\end{acknowledgments}

%\bibliographystyle{PRB}

%\bibliography{spinels}

%merlin.mbs apsrev4-1.bst 2010-07-25 4.21a (PWD, AO, DPC) hacked
%Control: key (0)
%Control: author (72) initials jnrlst
%Control: editor formatted (1) identically to author
%Control: production of article title (-1) disabled
%Control: page (0) single
%Control: year (1) truncated
%Control: production of eprint (0) enabled
%

\end{document}